\documentclass[12pt,a4paper]{article}
\usepackage{graphicx}
\usepackage[a4paper,left=1cm,right=1cm]{geometry}
\usepackage[latin1]{inputenc}
\usepackage{amsmath,amssymb,titling,authblk}
\usepackage{slashed}
\usepackage{amsfonts}
\usepackage{braket}
\usepackage{comment}
\usepackage{amsthm}
\usepackage{color}
\usepackage{dsfont}
\usepackage{slashed}
\usepackage{hyperref}
\usepackage{cite}
\usepackage{soul}

\allowdisplaybreaks

\newcommand{\del}{\partial}

\def\nn{\nonumber}
\newcommand{\be}{\begin{equation}}
\newcommand{\ee}{\end{equation}}
\newcommand{\bea}{\begin{eqnarray}}
\newcommand{\eea}{\end{eqnarray}}
\newcommand{\eqn}[1]{(\ref{#1})}
\def\beqa{\begin{eqnarray}}
\def\eeqa{\end{eqnarray}}
\newcommand{\R}{\mathbb{R}}

\begin{document}
\setlength{\droptitle}{-6pc}

\title{Double Quantization \vspace{5pt}}

\renewcommand\Affilfont{\itshape}
\setlength{\affilsep}{1.5em}
\renewcommand\Authands{ and }
 
\author[1,2]{Giulia Gubitosi\thanks{giulia.gubitosi@unina.it}}
\author[1,2,3]{Fedele Lizzi\thanks{fedele.lizzi@na.infn.it}}
\author[1,2,4]{Jos\'{e} Javier Relancio\thanks{relancio@unizar.es}}
\author[1,2]{Patrizia Vitale\thanks{patrizia.vitale@na.infn.it}}
\affil[1]{Dipartimento di Fisica ``Ettore Pancini'', Universit\`{a} di Napoli {\sl Federico II}, 80125 Naples, Italy\vspace{5pt}}
\affil[2]{INFN, Sezione di Napoli, 80125 Naples, Italy\vspace{5pt}}
\affil[3]{Departament de F\'isica Qu\`antica i Astrof\'isica and Institut de C\'encies del Cosmos (ICCUB),
Universitat de Barcelona, 08028 Barcelona, Spain\vspace{5pt}}
\affil[4]{Departamento de F\'{\i}sica Te\'orica and Centro de Astropart\'{\i}culas y F\'{\i}sica de Altas Energ\'{\i}as (CAPA),
Universidad de Zaragoza, Zaragoza 50009, Spain}

\date{}

\maketitle 

\begin{abstract}
In a quantum gravity theory, it is expected that the classical notion of spacetime disappears, leading to a quantum structure with new properties. A possible way to take into account these quantum effects is through a noncommutativity of spacetime coordinates. In the literature, there is not a clear way to describe at the same time a noncommutativity of spacetime and the phase-space noncommutativity of quantum mechanics. In this paper we address 
this issue by constructing a Drinfel'd twist in phase space which deals with both quantizations. This method can be applied to a noncommutativity which involves only space, leaving time aside. We apply our construction to the so-called $\lambda$-Minkowski and $\mathbb{R}^3_\lambda$ noncommutative spaces. 
\end{abstract}

\newpage

\newcommand{\verde}[1]{\textcolor[cmyk]{.83,.21,1,.08}{#1}}
\newcommand{\fedele}[1]{\verde{#1}}
\newcommand{\patrizia}{\textcolor{magenta}}

\section{Introduction}

A significant portion of research on quantum gravity aims at identifying suitable formalisms to describe the nontrivial properties of spacetime at the Planck scale. 
One of the most studied proposals -- and, in fact, the oldest~\cite{snyder} -- is that of spacetime noncommutativity. This  takes inspiration from the developments that lead to the formulation of quantum mechanics as a deformation of the  phase space of classical mechanics into a noncommutative phase space~\cite{Bayen:1977pr,Bayen:1977ha,Bayen:1977hb}.

In quantum mechanics, classical phase-space variables, $x^i,p_j$, are turned into non-commuting operators $\hat x^i,\hat p_j$, that satisfy the Heisenberg commutation relations, $\left[\hat x^i,\hat p_j\right]=i \hbar \delta^i_j$. This implies that the notion of a point in phase space is lost, due to the fundamental uncertainty relations $\Delta x^i\Delta p_i\geq \frac{\hbar}{2} $. The new fundamental constant, $\hbar$, with the dimensions of a length by a momentum, sets the amount of such uncertainty. 
In analogy to this procedure of quantization of phase space in quantum mechanics,  the proposal of spacetime noncommutativity is that spacetime coordinates, $x^\mu$, turn into non-commuting operators, $\hat x^\mu$, that satisfy the commutation relations $\left[\hat x^\mu,\hat x^\nu\right]=i \Theta^{\mu\nu}$. 

The quantity $\Theta^{\mu\nu}$ can be a constant matrix (in this case the noncommutativity is called \emph{canonical}) or  a function of the coordinates themselves, depending on the model\footnote{In the celebrated applications of noncommutative geometry started in~\cite{Doplicher:1994tu,Seiberg:1999vs} 
the quantity $\Theta=\theta$ was constant. In other interesting cases, with noncommutativity of Lie-algebra type,  $\Theta$ is  a linear function of  coordinates; this is the case of $\kappa$-Minkowski, or $\lambda$-Minkowski discussed in this paper.}. In any case, these commutation relations imply that one can no longer talk about points in spacetime, since coordinates are subject to the fundamental uncertainty $\Delta x^\mu\Delta x^\nu\geq \frac{\langle\Theta^{\mu\nu}\rangle}{2} $. The amount of such uncertainty is governed by some length scale $\lambda$, which plays a role analogous to that of the Planck constant in quantum mechanics. If spacetime noncommutativity is to be ascribed to quantum gravitational effects, then it would be natural to link this scale to the Planck length $l_P\equiv\sqrt{\hbar G/c^3}$. While this is a reasonable assumption, in principle the two noncommutativity constants, $\hbar$ and $\lambda$, are not necessarily related \cite{DiCasola:2014xya, Sorkin:2007qi} (see also \cite[Sec. 2.3]{Addazi:2021xuf} and references therein), and one could consider the two quantization procedures independently. \footnote{Two different quantization parameters appear also in  \cite{Ballesteros1995}, where a deformation of the Moyal-Weyl star-product that generates the quantization of phase space was considered, producing a quantum group deformation of the Heisenberg algebra governed by both $\hbar$ and a quantum deformation parameter.}

The goal of this work is to explore the relation between the quantization of phase space and that of spacetime. We will do that  by relying mostly on a specific quantization procedure and a specific noncommutative spacetime model, which are especially well suited to our scopes. 
Specifically, the quantization procedure we will adopt for both phase-space and spacetime quantization is based on \emph{deformation quantization}. The application of this approach to transition from classical to quantum mechanics  can be traced back to the early developments of quantum mechanics~\cite{Groenewold:1946kp, Moyal:1949sk}:  the commutative algebra of functions on phase space $(\vec x,\vec p)$  is deformed into a noncommutative algebra in which functions are multiplied using a \emph{noncommutative} product $\star_\hbar$, so that the star-commutator among  coordinate functions reproduces the desired Heisenberg commutator
\be
[x^i,p_j]_{\star_\hbar}\,\equiv\, x^i\star_\hbar p_j-p_j \star_\hbar x^i\,=\,i \hbar \delta^i_j \,. \label{commhbar}
\ee
There are different star-products which reproduce  Eq.~\eqref{commhbar}. They agree for coordinate functions, but not for generic functions. Their difference can be essentially understood in terms of ordering issues, and a choice must be made as to the particular form of the product. We will comment on this later.

In the spacetime canonical noncommutativity case, the analogue  of~\eqref{commhbar}  becomes
\be
[x^\mu,x^\nu]_{\star_\Theta}\,\equiv\, x^\mu\star_\Theta x^\nu-x^\nu \star_\Theta x^\mu\, =\,i\Theta^{\mu\nu}\,, \label{commtheta}
\ee
where the specific form of $\Theta$ depends on the noncommutative spacetime model considered, as we mentioned before. Also in this case the deformed product which for coordinate  functions reproduces  ~\eqref{commtheta} is not unique, and in the following we will fix its form following a standard prescription.

While formally very similar, the two quantizations defined, respectively, by Eqs. \eqref{commhbar} and \eqref{commtheta}, are conceptually very different.  We will call $\hbar$-quantization (or phase-space quantization) the usual procedure, leading to  quantum mechanics. This quantization is performed in the phase space of a single particle, in the framework of non-relativistic Heisenberg and Schr\"odinger mechanics. The appropriate symmetry for this framework is given by the Galilei group. 
Spacetime quantization, on the other hand,
focuses on the properties of spacetime itself. 
We  label  this quantization by the dimensionful parameter governing noncommutativity. Therefore, we talk of $\theta$-quantization, or~$\kappa-, \lambda-$, and so on. While the relativistic properties of the $\hbar$-quantization are straightforward,  it is easy to convince oneself that this is not the case for spacetime quantization. Indeed, \eqref{commtheta} is not symmetric under the action of the classical symmetries of spacetime, either Poincar\'e or Galilei.  Noncommutative geometries can nevertheless enjoy  deformed relativistic symmetries, implemented by a \emph{Hopf algebra}~\cite{Majid:1995qg}.

As we mentioned, spacetime quantization is usually performed ignoring $\hbar$, since its focus is to highlight the effects of nontrivial spacetime properties emerging at very small length scales, and $\hbar$ would become relevant only if one also considered  quantum matter.
However, it is necessary to include matter into the picture at some point, especially if one aims at doing phenomenological studies. It seems natural that this would require the presence of both constants -- the phase space noncommutativity constant $\hbar$ and the spacetime noncommutativity constant -- on the same footing, as well as a proper treatment of the symmetries of spacetime and matter. It is especially the symmetry aspects that make a \emph{double quantization}  challenging, since, as we have mentioned,  the quantization of spacetime generically calls for deformed symmetries, while standard quantization of phase space does not. In this paper, we will show that  double quantization is nevertheless possible and leads to a consistent treatment of symmetries. The main result we present is the construction of a spacetime and associated phase space in which  the algebra of functions of  positions and momenta is deformed via a $\star$-product which is covariant under a deformed Hopf algebra, compatibly with the deformation of the algebra of functions on spacetime.

Another issue that arises when attempting a simultaneous quantization of phase space and spacetime is the problem of time. If the latter is a non-commuting quantity realised in terms of a self-adjoint operator, then it will have  a real, continuous spectrum, ranging from $-\infty$ to $\infty$. But this will imply that the Hamiltonian, being its conjugate observable, should have as spectrum the entire real line as well.  Namely, the Hamiltonian would not  be bounded from below, leading to instability inconsistencies.  Nevertheless,  the bulk of the paper, and its examples, deal with a case in which time is a central element, commuting with all variables, thus circumventing this specific problem.

The main result of this paper is to develop a simultaneous quantization of phase space (related to quantum mechanics) and a quantum spacetime which we name\footnote{Alternative names have been used in the literature for this kind of angular noncommutativity, in which the central element is either a time or a space component. We retain the name $\varrho$-Minkowski for the case in which the central element is spatial, but will use $\lambda$-Minkowski in case the central element is time-like.}  $\lambda$-Minkowski~\cite{Gutt1983AnE,Gracia-Bondia:2001ynb,Lukierski:2005fc,Amelino-Camelia:2011ycm,Amelino-Camelia:2017pne,Ciric:2017rnf,DimitrijevicCiric:2018blz,Novikov:2019kit,Lizzi:2021dud}, for which the nontrivial spacetime commutators, of  ``angular kind'',   only involve  the spatial components. For this particular example, as we will see, it is possible to do such double quantization using a Drinfel'd twist. 

The structure of the paper is as follows. In Secs.~\ref{sec:lambdaM}-\ref{sec:twist} we review the noncommutativity of $\lambda$-Minkowski spacetime and its relation to  a specific  Drinfel'd twist. In Sec.~\ref{sec:twist_dq} we show that it is actually possible to define a twist operator for the whole phase space, which realises the sought for {\it double quantization}. Standard quantum mechanics is recovered in the limit $\lambda\rightarrow 0$ while spacetime quantization is obtained in the limit $\hbar\rightarrow 0$. In Secs.~\ref{sec:symmetries}-\ref{sec:hbarlambda}, we express the phase-space twist in terms of  generators of the Galilean group and   compute the coproducts of translation and rotation generators. The new twist depends both on  the quantum constant $\hbar$ and the parameter $\lambda$ related  to space  noncommutativity, but we find that it is always possible to take the limit in which one of the two goes to zero in a consistent way, thus reproducing either standard quantum mechanics, or spacetime quantization. In Sec.~\ref{sec:su2} we analyse the so-called  $\mathbb{R}^3_\lambda$ noncommutativity, namely a linear noncommutativity of $\mathfrak{su(2)}$ type \cite{Gracia-Bondia:2001ynb,Hammou:2001cc}, of which the $\lambda$-Minkowski Lie algebra is a kind of In\"ono\"u-Wigner contraction. We do find a double quantization in terms of a twist operator, but we find that it is not possible to disentangle space noncommutativity from the full phase-space one. Namely,  the limit $\hbar\rightarrow 0, \lambda$ finite, is ill defined for the coordinates commutators.   Finally, we summarise our results and discuss possible directions in Sec.~\ref{sec:conclusions}.


\section{\texorpdfstring{$\lambda$}{l}-Minkowski with phase-space quantization}
 \label{sec:lambdaM}

In order to  investigate the relation between phase-space and spacetime quantizations we  consider  a model where noncommutativity only affects spatial coordinates, in order to avoid,  in this exploratory work, having to deal with the  ambiguity added by  the role of  time  in the two constructions. Thus,  we will focus on specific  aspects inherent to the double quantization we are proposing and we will investigate the role of symmetries in this scenario.

The noncommutative spacetime model is characterised by the following relations:
\be
{}[ x^3,  x^1]\,=\, i \lambda   x^2\,, \qquad {}[ x^3,  x^2]\,=\, - i \lambda   x^1\,, \qquad  [x^1, x^2]= 0 \,, \qquad
{}[x^0,  x^i]\,=\,0\,.\label{commrel1}
\ee
We will call it $\lambda$-Minkowski in order to emphasize its similarity with the more famous $\kappa$-Minkowski~\cite{Lukierski:1991ff, Lukierski:1992dt, Majid:1994cy}, where a similar structure of the commutators holds between time and spatial coordinates. Moreover,   the model is related to $\R^3_\lambda$~\cite{Hammou:2001cc,Gracia-Bondia:2001ynb,Vitale:2012dz,Vitale:2014hca,
Kupriyanov:2015uxa,Juric:2016cfp}, a noncommutative spacetime with linear noncommutativity of $\mathfrak{su}(2)$ type, such that the algebra~\eqn{commrel1} is recovered by first considering a contraction of the $\mathfrak{su}(2)$ algebra and then adding a commutative time-coordinate (also see~\cite{Kupriyanov:2020sgx} where such a limit has been extensively studied in relation with non-commutative gauge theory and $L_\infty$ algebras). This model will be studied in Sec.~\ref{sec:su2}. 
Spaces with this kind of angular noncommutativity, which share with $\kappa$-Minkowski a Lie-algebra type noncommutativity, have been well studied, often in a variant, called $\varrho$-Minkowski, where $x^3$ and $x^0$ exchange their role~\cite{Chaichian:1998kp, Chaichian:2000ia,Lukierski:2005fc,Dolan:2006hv, Balachandran:2007sh, Meljanac:2006ui, Steinacker:2011wb,Amelino-Camelia:2011ycm,AmelinoCamelia:2011bm,DimitrijevicCiric:2018blz,Ciric:2019urb, Ciric:2019uab, Kurkov:2021kxa}.

 As we mentioned, spacetime noncommutativity is in principle independent of the usual phase-space noncommutativity governed by $\hbar$, and indeed the commutators \eqref{commrel1} only contain the spacetime noncommutativity parameter $\lambda$. This commutator can be understood in terms of a star product as in Eq.~\eqref{commtheta}, which only affects the algebra of functions on spacetime. If one is interested in studying the phenomenological consequences of~\eqref{commrel1}, it is however necessary to introduce matter into the picture, and with this comes the necessity to provide a compatible picture of phase space. In particular, one may ask whether the relations~\eqref{commrel1} are  compatible with a $\hbar$-quantized phase space, which would be  needed in order  to introduce quantum matter on such noncommutative spacetime.

 As in standard quantum mechanics, a semiclassical approximation of spacetime noncommutativity may be helpful to get an idea of the physical implications and of the mathematical structures. Therefore, spacetime commutators may  be replaced by non-trivial spacetime Poisson brackets and spacetime symmetries may be implemented as canonical transformations. We therefore preliminarily address the problem in a  semiclassical limit, where the spatial  commutators in~\eqn{commrel1} are replaced by Poisson brackets.
 \be\label{PBs}
 \{x^3, x^1\}\,=\,\bar\lambda x^2\,, \qquad   \{x^3, x^2\}\,=\,-\bar\lambda x^1\,, \qquad \{x^1,x^2\} \,=\, 0\,,
 \ee
 with $\bar\lambda$ directly proportional to $\lambda$ by some dimensional parameter, to be further specified.
 In such a context, the problem is well posed and amounts to look for extended Poisson brackets for the whole phase space, $T^*\R^3$,  which reduce to \eqn{PBs} on spacetime. Once a consistent Poisson algebra is obtained, we shall look for a  transformation in phase space, $(x,p)\rightarrow (\tilde x(x,p), \tilde p(x,p))$,  which maps the latter to the canonical Poisson bracket $\{\tilde x^i, \tilde p_j\}= \delta^i_j$. The quantization of this  structure is known. It is performed in the context of Weyl-Moyal quantization, with a well defined twist operator, which we shall transform back in terms of the original physical coordinates $(x^i, p_i)$. This new twist operator will provide the double quantization we are seeking. 
 
A generic Poisson bracket for the algebra of phase-space functions may be  associated with  a bivector field $\Lambda$ such that $\{f,g\}= \Lambda(df, dg)$. In the usual (canonical) case the structure is simply $\frac{\del}{\del x^i}\wedge \frac{\del}{\del p_i}$, where  $\wedge$ indicates the anti-symmetrized tensor product $a\wedge b=a\otimes b - b \otimes a$.
The search for a Poisson structure on $T^*\R^3$,  with a nice projection to the Poisson  tensor underlying \eqn{PBs}, say $\Pi$, amounts to a lift procedure (so called symplectic embedding\footnote{Here we follow an approach recently formalized in \cite{Kupriyanov:2021cws}, where by  symplectic embedding it is intended a specific generalization of the symplectic realization procedure due to Weinstein \cite{Weinstein,Crainic}. The relation among the two is discussed in some detail  in  \cite{Kurkov:2021kxa}. However,  in a wider  significance than here, symplectic and Poisson embeddings, as  lifting procedures,  are largely employed in the literature as useful tools in many different contexts.})  from space to phase space.
It can be  recast in the following form
 \be\label{poitenstra}
  \Pi\,=\, \Pi^{ij} \frac{\del}{\del x^i} \wedge  \frac{\del}{\del x^j}\rightarrow \Lambda\,=\,  \Pi^{ij} \frac{\del}{\del x^i} \wedge  \frac{\del}{\del x^j}- \gamma_i^{j}   \frac{\del}{\del x^i}\wedge \frac{\del}{\del p_j}\,.
 \ee
 We require that the lifted bracket be canonical in some new coordinates $(\tilde x(x,p), \tilde p(x,p) )$, namely that  $\Lambda= \frac{\del}{\del \tilde x^i} \wedge  \frac{\del}{\del \tilde p_i} $, and this imposes that  the matrix 
$\gamma(x,p)$  be the inverse of the Jacobian matrix $J= (\del \tilde x^i/\del x^j)$. 
Moreover, the requirement that Jacobi identity holds yields an equation for  $\gamma$,
 \be
\gamma_{m}^{n} \frac{\del}{\del p_m} \gamma^{k}_{\ell} - \gamma^{k}_{m} \frac{\del}{\del p_m}  \gamma^{n}_{\ell} 
+ \Pi^{n m} \frac{\del}{\del x^m}  \gamma^{k}_{\ell} 
- \Pi^{km} \frac{\del}{\del x^m}  \gamma^{n}_{\ell} 
- \gamma^{m}_{\ell}\frac{\del}{\del x^m} \Pi^{nk} \,=\, 0\,. \label{masterp1}
\ee
A solution, easily checked by a direct calculation, is 
\begin{align}
 \tilde x^1\,&=\,\ x^1\,,\ \tilde x^2\,=\, x^2\,,\ \tilde x^3\,=\, x^3 + \bar\lambda( x^2p_1-x^1 p_2)\,,\ \label{tras1}\\
 \tilde p_i &\,=\, p_i\,. \label{tras2}
\end{align}
When passing to commutators,  canonical quantization for the   algebra of phase-space coordinates reads
\begin{equation}
    [\tilde x^i, \tilde x^j]\,=\,0\,,\qquad[\tilde x^i,  \tilde p_j]\,=\,i \hbar \delta^i_j\,,\qquad  
    [ \tilde p_i, \tilde p_j]\,=\,0\,,
    \label{eq:Heisenberg_qm}
\end{equation}
with the time coordinate $\tilde x^0=x^0$ in the center of the algebra. 
This implies that, on choosing $\bar\lambda=\lambda/\hbar$, the ``true'' phase-space coordinates $x^i$ and $p_j$ satisfy
\be
{[}x^3,  x^i{]}\,= i\lambda  \epsilon^{3i}\,_k x^k, \qquad {[}x^1,  x^2{]}\,=0 \qquad {[}x^i,  p_j{]} \,=\, i\hbar \delta^i_j+ i \lambda \delta^{i3} {\epsilon_{3j}}^k  p_k\,, \qquad
{[} p_i,  p_j{]} \,=\, 0\,. \label{commrel3}
\ee
Therefore, standard quantum mechanics is recovered in the limit $\lambda\rightarrow 0$ while spacetime quantization is obtained in the limit $\hbar\rightarrow 0$.

Notice that the  choice \eqn{tras1}-\eqref{tras2} is not unique. Other solutions are possible. Also notice that Eq. \eqn{poitenstra} does not contain  terms of the form  $\del_{p_i}\wedge\del_{p_j}$, implying that the Poisson bracket of momenta  is chosen to be zero right from the beginning. \footnote{From a physical point of view, we notice that  $\del_{p_i}\wedge\del_{p_j}$ terms would induce nonvanishing commutators between spatial momenta, usually linked to nonvanishing spacetime curvature, which is not present in this context.} In the symplectic realization procedure \cite{Weinstein} this is a consequence of the formalism, whereas in the symplectic embedding approach it is a starting  assumption, a kind of minimal deformation.   Our choice could anyways be  relaxed, yielding to a generalization of Eq. \eqn{masterp1}, with a new family of solutions.  This choice has a nontrivial consequence, as we will see later, that terms which combine two derivatives with respect to the momentum will not appear in the twist.  

Clearly this observation is a good starting point, since we have produced a phase space whose commutation relations are compatible with those of spacetime (Eqs.~\eqref{commrel1} and \eqref{commrel3} close the Jacobi identities) and reduce to those of standard quantum mechanics in the limit $\lambda\rightarrow 0$. However, a few issues need to be addressed in order for the model defined by~\eqref{commrel1} and~\eqref{commrel3} to make sense.
First of all, while the commutators in~\eqref{eq:Heisenberg_qm} can be understood in terms of the star-product of quantum mechanics~\eqref{commhbar}, we still have to work out  the star product which gives rise to the commutators in~\eqref{commrel3}. 
Secondly, while the symmetries of an algebra of the kind of~\eqref{eq:Heisenberg_qm} are the Galilei symmetries, clearly this is not the case for the algebra defined by~\eqref{commrel1} and~\eqref{commrel3}.  
In the following we will show that  a fundamental aspect  in solving both of these issues  is that the  noncommutativity of~\eqref{commrel1} and~\eqref{commrel3} can be derived  from a unique \emph{Drinfel'd twist.}

\section{Drinfel'd Twists and Hopf Algebra}
\label{sec:twist}

  In this section we review the concept of twist and how it is used to deform the algebra of smooth functions $C^\infty(M)$
  on spacetime $M$, together with the associated symmetries. 
  
  The twist operator $\mathcal{F}$ is defined in terms of elements of the tensor product $U(\mathfrak{g})\otimes U(\mathfrak{g})$, with $\mathfrak{g}$ some Lie algebra acting on the algebra of functions $C^\infty(M)$, and $U(\mathfrak{g})$ its universal enveloping algebra. Given an invertible $\mathcal{F}\in U(\mathfrak{g})\otimes U(\mathfrak{g})$, it is possible to define a noncommutative star product in $C^\infty(M)$, which is  associative, provided $\mathcal{F}$ satisfies the cocycle condition
  \be
  (\mathcal{F}\otimes 1)(\Delta\otimes id) \mathcal{F}\,=\, (1\otimes \mathcal{F})(id \otimes \Delta) \mathcal{F}\,, \label{cocycle}
  \ee
  with $\Delta: X\in \mathfrak{g}\rightarrow X\otimes 1+ 1\otimes X $ the primitive coproduct for the Lie algebra. Given $\mu:f\otimes g\rightarrow f\cdot g$ the standard pointwise multiplication in $C^\infty(M)$, the deformed product of functions is then defined according to
  \be
  f\star g\,:= \,\mu \circ \mathcal{F}^{-1}(f\otimes g)\,.
  \ee
  The twist is further supplemented with the normalization condition
  \be
  \mu(\epsilon\otimes 1) \mathcal{F}= \mu (1\otimes \epsilon) \mathcal{F}\,=\,1\,,
  \ee
  where $\epsilon$ is the co-unit. 
  
  A special family of twist operators is represented by the Abelian twists, so named because they have support in an Abelian algebra. Among them, in the next section we shall make use of  the Moyal-Weyl twist of quantum mechanics, which is defined  in terms of  translation generators.  
  But also the algebra of $\lambda$-Minkowski  spacetime \eqn{commrel1}, when    realised in terms of star commutators, is  associated with an  Abelian twist~\cite{Majid:1995qg}:
  \be
  \mathcal{F}\,=\, \exp \left\{ -\frac{i\lambda}{2}\left[\del_3\otimes(x^2\del_1-x^1\del_2)-(x^2\del_1-x^1\del_2) \otimes \del_3 \right]\right\}\,.
   \label{NCSTtwist}
  \ee
The search for a consistent double quantization which reduces to the latter when $\hbar\rightarrow 0$ is the object of    the forthcoming sections. 

Before proceeding further, let us first recall that
the existence of a well-behaved twist operator allows for the twisting of any other geometric structure (see for example~\cite[Chapt.~7]{aschieri2009noncommutative}) related to the quantum spacetime under analysis. We briefly sketch how the procedure applies to symmetries. 

 In the twist approach, two slightly different quantum Hopf algebras of symmetries may be defined, indicated in~\cite[Chapt.~7]{aschieri2009noncommutative} by $\mathcal{U}_{\mathcal{F}}(\mathfrak{g})$ and $\mathcal{U}_\star(\mathfrak{g})$, which are however isomorphic. Here we shall work  with  $\mathcal{U}_{\mathcal{F}}(\mathfrak{g})$, while we refer to  the cited literature for details on the alternative construction. As for $\mathcal{U}_{\mathcal{F}}(\mathfrak{g})$,  the Lie algebra generators act undeformed on a single copy of the algebra of observables, with standard Lie brackets. However, in order to act on  products of observables (and therefore on the star-commutators which are associated with ~\eqn{commrel1}), the coproduct of   Lie algebra generators  has to be twisted for the consistency of the whole quantum enveloping algebra, according to 
 \be\Delta_{\mathcal{F}}\,=\, \mathcal{F}\Delta\mathcal{F}^{-1}\,. 
 \ee
 This entails  a twisted   Leibniz rule,   so that   for $X\in \mathfrak{g}$, $f, g \in C^\infty(M)$
\be
X\triangleright f\star g\,:=\, \mu_\star\circ\Delta_{\mathcal{F}}(X)(f\otimes g)\,.
\ee
We shall exhibit  in the forthcoming sections an explicit realisation for the twisted coproducts of the generators of the Galilei relativity group associated with the noncommutative spaces of interest.

\section{The Twist for double quantization}
\label{sec:twist_dq}

In the previous section we have reviewed how the Drinfel'd twist applies to the algebra of functions on spacetime  in order to turn it into a noncommutative algebra and work out the corresponding relativistic symmetries as Hopf algebra deformations of the classical ones.

In fact, a twist procedure can also be used to quantize phase space into the one of quantum mechanics. 
Specifically, the  $\hbar$-quantization producing the commutation relations \eqref{eq:Heisenberg_qm} is induced from the following twist acting on the algebra of functions on phase space:
\be
\mathcal{F}_\hbar\,=\, \exp\left[-\frac{i\hbar}{2}\left( \frac{\del}{\del \tilde x^i}\wedge \frac{\del}{\del  \tilde p_i}\right)\right]\,, \label{QMtwist}
\ee

This is an Abelian twist, since $\frac{\del}{\del \tilde x^i}$ commutes with $\frac{\del}{\del \tilde p_j}$, and as such it satisfies the cocycle conditions. 

In order to produce a doubly-quantized model, with both  phase-space noncommutativity, governed by $\hbar$, and  spacetime noncommutativity, governed by $\lambda$, one cannot simply apply the two twists, \eqref{NCSTtwist} and \eqref{QMtwist}, one after the other, since they would not be acting on the correct algebra of functions. {The solution is represented by the symplectic embedding worked out in Sec.~\ref{sec:lambdaM}, which allows for expressing the Poisson tensor  of $\lambda$-deformed spacetime as a projection of a Poisson tensor on the full phase space, according to the mapping \eqn{poitenstra}. The latter  is in turn mapped to the canonical  Poisson tensor of phase space, $P= \frac{\del}{\del \tilde x^i}\wedge \frac{\del}{\del \tilde p_i}$,   through the non-linear transformations \eqn{tras1}-\eqn{tras2}. This procedure can be extended to the quantum setting, where we will show that  it is possible to obtain  a new twist operator comprising both quantizations at once, by simply exponentiating  the phase-space Poisson tensors described above. Roughly speaking, this can be worked out by starting from the $\hbar$-twist \eqref{QMtwist} and applying to it the change of variables \eqn{tras1}-\eqn{tras2} 
\be
\frac{\del}{\del \tilde x^i}\,=\,\frac{\del}{\del   x^i} +\frac{\lambda}{\hbar}{\epsilon_i}^{j3}  p_j \frac{\del}{\del   x^3} \,,\qquad
\frac{\del}{\del \tilde p_i}\,=\,\frac{\del}{\del   p_i} -\frac{\lambda}{\hbar}{\epsilon^i}_{j3}   x^j \frac{\del}{\del  x^3}\,. 
\label{eq:r3partials}
\ee}
With this change of variables the Moyal twist~\eqref{QMtwist}  of standard quantum mechanics becomes, \beqa
\mathcal{F}&=& \exp\left[-\frac{i\hbar}{2} \left( \frac{\del}{\del  x^i}\wedge \frac{\del}{\del  p_i}\right)
+ {\frac{i\lambda}{2}\left( \left( x^1\frac{\del}{\del  x^3} \wedge \frac{\del}{\del   x^2}-  x^2\frac{\del}{\del  x^3}\wedge \frac{\del}{ \del  x^1}\right) 
+  \left ( p_1\frac{\del}{\del  x^3} \wedge  \frac{\del}{\del  p_2}- p_2\frac{\del}{\del  x^3} \wedge \frac{\del}{\del  p_1}\right)\right) }\right. \nonumber\\
 &&+\left.
{ i\frac{\lambda^2}{2\hbar} \left(p_1\frac{\del}{\del x^3}\wedge x^1\frac{\del}{\del x^3}+p_2\frac{\del}{\del x^3}\wedge x^2\frac{\del}{\del x^3} 
 \right)}
\right]\,.
 \label{newtwist}
\eeqa
This twist provides the double quantization. It is worth noting that there are no terms of the form $\partial/\partial p_i \wedge \partial/\partial p_j$. This is due to the fact that $\tilde p_i = p_i$, so when computing~\eqref{eq:r3partials},  terms proportional to $\partial/\partial p_i $ do not appear in $\partial/\partial \tilde x^i$. 

It is now possible to see the following limits of the noncommutativity constants. The limit $\lambda\to0$ gives quantum mechanics with standard, i.e., commutative,  configuration space. The limit $\hbar\to0$ gives the $\lambda$-Minkowski space we started with\footnote{{The last term will not contribute to the commutators of the coordinates, nor (in the limit of small $\hbar$)  to the product of general rapid decreasing functions because it is rapidly oscillating.}} \emph{with a caveat}. The twist \eqref{newtwist} gives nontrivial commutations between $x^3$ and $p_i$, analogous to the ones between the $x$'s:
\be
[x^3,p_1]\,=\, -i\lambda p_2,\qquad [x^3,p_2]=i\lambda p_1\,. \label{newxp}
\ee
These terms are necessary for phase-space coordinates to close a Lie algebra. Indeed, starting form Eq.~\eqref{commrel1} and computing its commutator with momenta, one has 
\be
[[x^3,x^1],p_2]\,=\,-i[\lambda x^2,p_2]=\lambda\hbar\,.
\ee
A simple calculation shows that, in order  to satisfy Jacobi identity  one has to impose~\eqref{newxp}. In the $\hbar\to0$ limit, while the commutators \eqref{newxp} survive,  one can be content with just studying the spacetime noncommutativity, consistently ignore the commutators \eqref{newxp} and reduce to the $\lambda$-Minkowski spacetime model we started with.  The commutators \eqref{newxp} are only needed if one wants to study the full phase space associated to $\lambda$-Minkowski spacetime. They appear in the $\hbar\rightarrow 0$ limit because we are taking the limit on the full phase space, but reduction to spacetime can   be done consistently.

The new twist \eqn{newtwist} provides the sought-for double quantization of both phase space and spacetime, with spacetime noncommutativity ruled by Eq.~\eqn{commrel1} and star product expressed in terms of the twist as usually done:
\be
(f\star g) ( x,  p) \,=\, \mathcal{F} \triangleright(f, g)\,. 
\ee

As expected, given our comments earlier in this section, the twist \eqn{newtwist} cannot be written as the product of two twists, one only involving $\hbar$ and the other only $\lambda$, but  we can separately consider the limits of standard quantum mechanics, $\lambda\rightarrow 0$, and of pure spacetime noncommutativity, $\hbar\rightarrow 0$.

Having defined a consistent $\star$-product compatible with the double quantization defined by the commutation relations \eqref{commrel1} and \eqref{commrel3}, so that these commutators are  well-defined, the issue remains of establishing whether such model is compatible with some deformation of the standard relativistic symmetries of quantum mechanics, namely of the Galilean symmetries.  This aspect is addressed in the following section.

\section{Symmetries of the doubly-quantized phase space}
\label{sec:symmetries}

As we discussed in Sec \ref{sec:twist}, the twist that turns a commutative algebra of functions into a noncommutative one also affects the symmetries of this algebra, turning a classical algebra of symmetries into a Hopf algebra.

Before going to the doubly-quantized phase space defined by the twist \eqref{newtwist}, let us review how standard quantization   is performed within a twisting procedure. The noncommutative phase space of quantum mechanics results from  Moyal-Weyl quantization (via the twist \eqref{QMtwist}) of the classical non-relativistic phase space, that enjoys Galilean symmetries generated by the time translation $P_0$, space translations $P_i$, rotations $J_i$, and Galilean boosts $K_i$:
\be\label{Galilei}
\begin{array}{lll}
\left[J_i,J_j\right]\,=\,i\epsilon_{ijk}J_k\,,\qquad&\left[J_i,P_j\right]\,=\,i\epsilon_{ijk}P_k\,, \qquad  &\left[J_i,K_j\right]\,=\,i\epsilon_{ijk}K_k\,,\\
 \left[K_i,K_j\right]\,=\,0\,,   & \left[K_i,P_j\right]\,=\,0\,, &  \left[K_i,P_0\right]\,=\,i P_i\,,\\
  \left[P_0,P_i\right]\,=\,0\,,    &   \left[P_i,P_j\right]\,=\,0\,, &   \left[P_0,J_i\right]\,=\,0\,. \\
\end{array}
\ee 
Then the twist \eqref{QMtwist}  should be applied to the Galilei algebra in order to find the symmetries of standard (non-relativistic) quantum mechanics. It turns out that such twist acts trivially on  the  Galilei algebra, leaving the coproducts of its generators  unaffected. Namely, as is well-known, quantum non-relativistic symmetries  are described by the same undeformed Galilei algebra~\cite{LevyLeblond:1974zj} as classical mechanics. 
To see this explicitly, we recall that the  twist procedure acts on the coproducts of the generators $G$ of the algebra according to:
\be
\Delta^{ \mathcal{F}}G\,=\, \mathcal{F}\circ \Delta^{(0)}G\circ  \mathcal{F}^{-1} \,,
\ee
while leaving the commutators invariant.
In order to explicitly compute the coproducts, we need to express the twist \eqref{QMtwist} in terms of  the generators of the Galilei algebra.

The standard procedure is to realise the infinitesimal generators  as differential operators in configuration space and perform the so-called cotangent lift~ (see for example \cite{abraham}): 
\be
G\,=\, f^i(x) \frac{\del}{\del x^i} \longrightarrow G^C\,=\, f^i(x,t) \frac{\del}{\del x^i}  + \frac{d f^j}{d t}\delta_{ji } \frac{\del}{\del p_i} \,,  
\ee
where the second term becomes  $ = -  p_j  \frac{\del f^j}{\del x^i} \frac{\del}{\del p_i}$ for $t$-independent generators (which {\it is not} the case for the boosts, $K_i$).
Through the cotangent lift, the generators of the Galilei algebra become:
\be\label{cotanlift}
\begin{array}{llllll}
 P_{i}\,=\, \frac{\del}{\del x^i}&& \longrightarrow && P^C_{i}\,=\, \frac{\del}{\del x^i}\\
 J_{i}\,=\, {\epsilon_{ij}}^k x^j \frac{\del}{\del x^k} &&\longrightarrow&& J^C_{i}\,=\, {\epsilon_{ij}}^k x^j \frac{\del}{\del x^k}+ {\epsilon_{ki}}^{j} p_j \frac{\del}{\del p_k}\\
K_{i}\,=\, t  \frac{\del}{\del x^i}&& \longrightarrow&& K^C_{i}\,=\, t  \frac{\del}{\del x^i} + m   \frac{\del}{\del p_i} \\
 P_0\,=\, \frac{\del}{\del t} && \longrightarrow&& P_0^C \,=\, \frac{\del}{\del t}\end{array}
\ee 
 where we have inserted a mass scale for the boosts generators for dimensional reasons. Also notice  that time has a special role in non-relativistic mechanics, meaning that we have to extend the phase space $T^* M \rightarrow T^* M \times R$ (and $t$ has no conjugate variable). By construction, the lifted generators obey the same algebra as the un-lifted ones. 
 
From here we see that 
\be\label{representation}
\frac{\del}{\del x^i}\,=\, P^C_{i}\,, \qquad
\frac{\del}{\del p_i}\,=\, \frac{1}{m } K^C_{i} - \frac{t}{m} P^C_{i}\,,
\ee
and the twist of quantum mechanics \eqref{QMtwist} can be written as: 
\be
\mathcal{F}_\hbar\,=\, \exp\left[-i\frac{\hbar}{2m } \left(P^C_{i} \wedge K^C_{i}  \right)\right]\,.
\ee

With this representation of the twist at hand we can now compute the coproducts and find that  they are undeformed, as expected. As an example, we compute  the coproduct $\Delta(P_0)$:
\be
\Delta^{\mathcal F_\hbar}(P_0)\,=\, P_0 \otimes 1+ 1\otimes P_0 + \frac{i\hbar}{m}\left(-P_i \otimes [P_0, K_i]+ [P_0, K_i]\otimes P_i\right)+O(\hbar^2)
\,=\, P_0 \otimes 1+ 1\otimes P_0 \,,
\ee
where the superscript $C$ for the lifted generators is understood.
It is easy to see that there is no  modification of the coproduct of $P_0$ coming from  higher orders in $\hbar$. This is due to fact that $[P_0,P_i]=0$. 
Similar calculations for all generators of the Galilei algebra can be performed, yielding undeformed co-products for all of them. Namely,  the symmetries of standard quantum mechanics are undeformed in the twist approach, as expected.

Having seen how the twist allows to correctly deduce the symmetries of the noncommutative phase space of quantum mechanics starting from those of the commutative phase space of Newtonian mechanics, we can now apply a similar procedure in order to find the symmetries associated to the doubly-quantized phase space which is generated by the twist \eqref{newtwist}. In this case, we should no longer act on the algebra of symmetries of Newtonian mechanics  with the twist \eqref{QMtwist}, but with the new twist \eqref{newtwist}, that quantizes  phase space and spacetime  at once.

Following the same steps as before, by using the relations \eqref{representation} we write the {first order of the } complete twist \eqn{newtwist} in terms of the cotangent lift of generators of the Galilei algebra:
\be\label{eq:Twist}
{\mathcal{F}^{(1)}\,=\, \mathbf{1}\otimes \mathbf{1}-\frac{i\hbar}{2m } \left( P^C_{i} \wedge  K^C_{i}   \right)- \frac{i\lambda}{2}\left(J^C_3\wedge P^C_3 \right){+O(\lambda)^2} \,.}
\ee

We can now show that already at the first order this twist induces a modification of the coproduct of the Galilei generators, clearly only dependent on the spacetime noncommutativity parameter $\lambda$ (as  before,   the superscript of the cotangent lift is understood from now on): 
\beqa
\Delta^{\mathcal F}(P_0)&=& P_0 \otimes 1+ 1\otimes P_0\,, \nonumber\\
\Delta^{\mathcal F}(P_j)&=& P_j \otimes 1+ 1\otimes P_j+\frac{\lambda}{2}\epsilon_{3j}\,^{k}P_k\wedge P_3\,,\nonumber\\
\Delta^{\mathcal F}(K_j)&=& K_j \otimes 1+ 1\otimes K_j+\frac{\lambda}{2}\epsilon_{3j}\,^{k}K_k\wedge P_3\,,\nonumber\\
\Delta^{\mathcal F}(J_j)&=& J_j \otimes 1+ 1\otimes J_j+\frac{\lambda}{2}\epsilon_{3j}\,^{k}(J_k\wedge P_3+ J_3\wedge P_k)\,.
\eeqa
 All the  generators are affected by space noncommutativity, with two notable exceptions: the time translation generator, $P_0$, which is undeformed as expected since we are in a non-relativistic setting, and one of the rotation generators, $J_3$. This is also understandable given the structure of the commutator between spatial coordinates, Eq.~\eqref{commrel1}: if we rotate around the  $\hat 3$ direction we are mixing the $\hat 1$ and $\hat 2$ directions, which commute among themselves, so this rotation is not affected by spatial noncommutativity.  Notice that to the first order, these coproducts reproduce the ones that are found when considering noncommutativity of spacetime alone as done in~\cite{DimitrijevicCiric:2018blz}. In the following we will show why this is the case.

\section{Interplay between \texorpdfstring{$\hbar$}{hb} and \texorpdfstring{$\lambda$}{l}} \label{sec:hbarlambda}

In order to see the effects of the interplay between the phase-space noncommutativity parameter $\hbar$ and the spacetime noncommutativity parameter $\lambda$, we need to compute the twist of the coproducts {including mixed terms proportional to $\hbar\lambda$. 
To this aim, it is useful to observe that, for a generator $G$ of the Galilei algebra and writing the twist schematically as $\mathcal F=\exp({X\otimes Y})$, with $X$ and $Y$  Galilei generators as well, the twist acts as follows:}
\be
\Delta^{\mathcal F}(G)\,\simeq\, \Delta^{(0)}(G)+\left[ (X\otimes Y), \Delta^{(0)}(G)\right]+\frac{1}{2}\left[(X\otimes Y),\left[( X\otimes Y), \Delta^{(0)}(G)\right]\right]\,,\label{eq:twistexpansion}
\ee
where $\Delta^{(0)}(G)$ is the zero-order expansion of the coproduct, which coincides with the primitive coproduct for $G$.
 By specializing to the twist \eqref{eq:Twist} one has: 
\beqa
X\otimes Y\,=\,-\frac{i\hbar}{2m }\left(  P_{i} \wedge K_{i}\right) -\frac{i\lambda}{2} \left(  J_3\wedge P_3  \right)\,,
\eeqa
so the mixed terms in $\hbar\lambda$ in the coproduct $\Delta^{\mathcal F}(G)$ can only be generated by the last term in the expansion \eqref{eq:twistexpansion}. The inner commutator in this term only gets contributions from the $\lambda$ part of the twist, for any generator $G$ (this explains why the coproducts are not modified by the $\hbar$ part of the twist alone). So, in order to get the mixed terms, one only needs  to use the $\hbar$ part of the twist for the outer commutator. In other words, mixed $\hbar\lambda$ contributions in the coproduct of $G$ come from terms of the form:
\be
\left[\frac{i \hbar}{2 m } \left( P_{i} \wedge K_{i} \right) ,\left[\frac{i\lambda}{2} \left( J_3\wedge P_3\right)  , \Delta^{(0)}(G)\right]\right]\,.
\ee
By inspecting this object it is straightforward to see that mixed terms  only emerge for $G=J_1$ or $G=J_2$, because the generators $P_i$ and $K_j$ commute among themselves. So only the rotation sector is affected by mixed terms in $\hbar\lambda$, while the translation and boost sectors are only affected by the spacetime noncommutativity parameter $\lambda$.

In summary, we find that the 
{the modification of}  the coproducts of the Galilei generators is: 
\begin{eqnarray}
\Delta^{\mathcal F}(P_0)&=& P_0 \otimes 1+ 1\otimes P_0\,, \nonumber\\
\Delta^{\mathcal F}(P_j)&=&P_j \otimes 1+ 1\otimes P_j+\frac{\lambda}{2}\epsilon_{3j}\,^{k}P_k\wedge P_3\,,\nonumber\\
\Delta^{\mathcal F}(K_j)&=&K_j \otimes 1+ 1\otimes K_j+\frac{\lambda}{2}\epsilon_{3j}\,^{k} K_k\wedge P_3\,,\nonumber\\
\Delta^{\mathcal F}(J_1)&=& J_1 \otimes 1+ 1\otimes J_1+\frac{\lambda}{2} \left( J_2 \wedge P_3+J_3 \wedge P_2\right)\nonumber\\
&&+\frac{\lambda \hbar}{8 m} \left( {\epsilon^{ij}}_3 \left( P_i \otimes K_j- K_i \otimes P_j\right) P_2+{\epsilon^{ij}}_2 \left( P_i \otimes K_j- K_i \otimes P_j\right) P_3\right.\nonumber\\
&&+ \left.{\epsilon^{ij}}_3 P_2  \left( P_i \otimes K_j- K_i \otimes P_j\right)+{\epsilon^{ij}}_2 P_3\left( P_i \otimes K_j- K_i \otimes P_j\right)\right)\nonumber\\
\Delta^{\mathcal F}(J_2)&=& J_2 \otimes 1+ 1\otimes J_2-\frac{\lambda}{2} \left( J_1 \wedge P_3+J_3 \wedge P_1\right)\nonumber\\
&&-\frac{\lambda \hbar}{8 m} \left( {\epsilon^{ij}}_3 \left( P_i \otimes K_j- K_i \otimes P_j\right) P_1+{\epsilon^{ij}}_1 \left( P_i \otimes K_j- K_i \otimes P_j\right) P_3\right.\nonumber\\
&&+ \left.{\epsilon^{ij}}_3 P_1 \left( P_i \otimes K_j- K_i \otimes P_j\right)+{\epsilon^{ij}}_1 P_3 \left( P_i \otimes K_j- K_i \otimes P_j\right)\right)\nonumber\\
\Delta^{\mathcal F}(J_3)&=& J_3 \otimes 1+ 1\otimes J_3\,,
\end{eqnarray}
As we mentioned, only the rotation generators are affected by the interplay between spacetime and phase-space noncommutativity. We see, however, that such interplay does not spoil the trivial coproduct of the generator of rotations around the $\hat 3$ direction. Again, this has an intuitive explanation, since the phase-space noncommutativity does not spoil the fact that rotations around the $\hat 3$ direction mix the directions $\hat 1$ and $\hat 2$, which are commutative. In fact, looking at the phase-space commutator~\eqref{commrel3}, we see that $[x^1,p_2]=[x^2,p_1]=0$.

\section{\texorpdfstring{$\mathbb{R}^3_\lambda$}{R3t} noncommutativity}
\label{sec:su2}

So far we have studied as a main example the double quantization of the phase space that is compatible with $\lambda$-Minkowski noncommutative  spacetime. In this example, the limit in which either of the two dimensionful constants ($\hbar$ and $\lambda$) vanishes can be performed independently. One may wonder if this is a special feature of $\lambda$-Minkowski, or it is a general property of noncommutative spacetime models. It turns out that the former hypotesis is correct, and the separability between spacetime and phase-space quantizations is a feature in general not shared by other models. 
In this section we provide an example of this case. Specifically, we start from another noncommutative space, quite well  studied in the literature, known as $\mathbb{R}^3_\lambda$ space~\cite{Hammou:2001cc,Gracia-Bondia:2001ynb,Vitale:2012dz,Vitale:2014hca}\footnote{The use of the same symbol for the noncommutative parameter is not casual. Indeed, as a Lie algebra,  $\lambda$-Minkowski may be obtained by the algebra of coordinates of $\R^3_\lambda$ as a contraction. See for example \cite{Gracia-Bondia:2001ynb} for details.  }. The commutators for the spatial coordinates are
\begin{equation}\label{$k$-Minkowski}
    [x^i,x^j]\,=\,i 2 \lambda {\epsilon^{ij}}_k x^k\,.
\end{equation}

In order to construct the double quantization of the phase space associated to the space noncommutativity \eqref{$k$-Minkowski}, we can proceed as for the $\lambda$-Minkowski case and express the noncommutative space coordinates as a function of the canonical phase-space ones, given in Eq.~\eqref{eq:Heisenberg_qm}. A particularly  simple solution is
\begin{equation}
    x^i\,=\,\tilde{x}^i-\frac{\lambda}{\hbar } {\epsilon^{ij}}_k \tilde{p}_j \tilde{x}^k+\frac{\lambda^2}{\hbar^2} \delta^{ij}\tilde{p}_j \tilde{p}_l \tilde{x}^l\,,
    \label{eq:change_x_su2}
\end{equation}
corresponding to the case in which $h=\lambda$ and $f=1$ for the general solution obtained in~\cite{Juric:2017bpr}. Notice that, unlike the $\lambda$-Minkowski case, Eq.~\eqref{tras1}, the relation between $x^i$ and $\tilde x^i$, is nonlinear in the momenta. Still, this relation may be inverted for $\tilde x^i$, yielding 
\begin{equation}
    \tilde{x}^i\,=\,\alpha(p) \left(x^i+\frac{\lambda}{\hbar} {\epsilon^{ij}}_k p_j x^k\right)\,,
\end{equation}
where 
\begin{equation}
  \alpha(p)\,=\, \left(1+\frac{\lambda^2}{\hbar^2 }p_l{\delta}^{l m} p_m\right)^{-1}\,.
\end{equation}
From the previous expression we can read the commutators in phase space
\begin{equation}\label{eq:RlambdaPS}
    [x^i,p_j]\,=\,i\left(\hbar \delta^i_j-  \lambda {\epsilon^{ik}}_j p_k+\frac{\lambda^2}{\hbar} \delta^{il}{p}_l {p}_j\right)\,,
\end{equation}
where again we notice the presence of a quadratic term in the momenta.
From the change of variables \eqref{eq:change_x_su2} we get  
\beqa
\frac{\del}{\del \tilde x^i} &=&\frac{\del}{\del   x^i}+\left(- \frac{\lambda}{\hbar}{\epsilon_i}^{jl}  p_j +\frac{\lambda^2}{\hbar^2} p_i p_j \delta^{jl} \right) \frac{\del}{\del   x^l}\,,\nn\\
\frac{\del}{\del \tilde p_i}&=&\frac{\del}{\del   p_i}+\left( \frac{\lambda}{\hbar}{\epsilon_k}^{i{l}}   \tilde{x}^k +\frac{\lambda^2}{\hbar^2}\left(\delta^{i l} p_m \tilde{x}^m+\delta^{lm}p_m \tilde{x}^i\right)\right) \frac{\del}{\del   x^l}\,, 
\eeqa
so that  the Moyal twist~\eqref{QMtwist} takes the form   
\begin{eqnarray}
\mathcal{F}&=& \exp\left[-\frac{i\hbar}{2} \left( \frac{\del}{\del  x^i}\wedge \frac{\del}{\del  p_i}\right)
+ \frac{i\lambda}{2} \left( {\epsilon^{ij}}_k p_j \frac{\del}{\del  x^i} \wedge \frac{\del}{\del  p_k}- {\epsilon^{ij}}_k \alpha(p) x^k  \frac{\del}{\del  x^i} \wedge \frac{\del}{\del  x^j}\right)  \right.\nonumber\\
&&\left.-\frac{i\lambda^2}{2\hbar}\left(p_l p_j \delta^{il}\frac{\del}{\del  x^i} \wedge \frac{\del}{\del  p_j}
{
+\frac{\del}{\del x^k}\wedge\left(p_mx^m\delta^{lk}+x^k p_m\delta^{lm}\right)\frac{\del}{\del x^l}
-x^k\frac{\del}{\del x^m}\wedge\left(p_k\frac{\del}{\del x^l}\delta^{lm}-p_l\delta^{lm}\frac{\del}{\del x^k}\right)
} 
\right)\right.
 \nonumber\\
 &&\left.{+O(\lambda^3)}\right]\,.
 \label{newtwistsu2}
\end{eqnarray}
As in the previous studied case, there are no terms of the form $\partial/\partial p_i \wedge \partial/\partial p_j$ since $\tilde p_i = p_i$.

The latter represents the double quantization of the $\R^3_\lambda$ algebra. 
Although this expression is not particularly transparent, we can analyse its possible limits. The limit $\lambda\rightarrow 0$
 is well defined and yields back standard phase-space quantization. 
{  The limit $\hbar\rightarrow 0$, instead,  yields divergences in the commutator \eqn{eq:RlambdaPS}.}
{Therefore, it is not possible to isolate $\lambda$-quantization from $\hbar$-quantization at the level of the commutators, since the $\hbar\rightarrow 0$ limit can only be taken together with the $\lambda\rightarrow 0$ limit, in such a way that $\lambda/\hbar$ does not diverge. 
 This represents 
 a crucial difference  with respect to the findings of   previous section for  $\lambda$-noncommutativity, Eq.~\eqref{newtwist}. We also notice that the latter, although being a contraction of $\R^3_\lambda$ at the Lie algebra level, is not related to $\R^3_\lambda$ as for their double quantization in terms of twists.

\section{Conclusions and outlook}
\label{sec:conclusions}

The goal of this work was to investigate whether spacetime and phase-space noncommutativity are mutually compatible and lead to models that enjoy relativistic symmetries, described by some deformation of the symmetry group of quantum mechanics, that of Galilei. We have done so within the formalism of deformation quantization, that is known to be fit to describe the two kinds of noncommutativity when taken on their own. Specifically, we have relied on the tools provided by the Drinfel'd twist and investigated whether it is possible to construct a twist that produces this double quantization, in such a way that when spacetime noncommutativity is turned off one recovers the standard quantization of quantum mechanics, governed by $\hbar$. 

As starting point for the double quantization, we have considered two different noncommutative spacetimes, $\lambda$-Minkowski and $\mathbb{R}^3_\lambda$, both characterized by a commuting time coordinate, but differing in the properties of the spatial commutators. 
For the first model, we have found that  the twist procedure leads to a deformation of the algebra of functions over phase space and spacetime which represents a truly ``quantum noncommutative geometry''. This construction admits a description of the relativistic symmetries in terms of a Hopf-algebra deformation of the Galilei algebra. In this algebra, all generators except the time translations have deformed coproducts, as in the $\lambda$-Minkowski noncommutative spacetime model. Moreover, coproducts of angular momenta contain new terms, involving the product of phase-space and spacetime noncommutativity constants, $\hbar$ and $\lambda$ respectively,  in addition to the usual coproduct one would have when only spacetime noncommutativity is turned on. Finally, we have shown that in this case all the structures built with the double quantization procedure  reduce to the standard ones of quantum mechanics (respectively, noncommutative spacetime), in the vanishing $\lambda$ (respectively, $\hbar$) limit.\footnote{The classical limit of quantum mechanics, where commutators are replaced by Poisson brackets, can be obtained by first performing the $\lambda\to 0$ limit, so to get vanishing spatial commutators, and then taking the limit of the phase space commutators as usual, $\frac{1}{i\hbar} [\cdot,\cdot]\xrightarrow{\hbar \to 0} \{\cdot,\cdot\}$.}
In contrast with the first scenario, for the second one involving the   $\mathbb{R}^3_\lambda$ spacetime we have found that the twist producing the double quantization does not allow for a smooth limit to the pure spacetime noncommutativity case.  This is due to the fact that the algebra $\mathbb{R}^3_\lambda$  does not admit a simple representation of the noncommutative coordinates as  functions of the commutative ones.

The relevance of the results found for the first model is made more apparent by the fact that we found examples, such as the second model we considered, where such a construction does not work.  Within the framework we considered here, the $\mathbb{R}^3_\lambda$ can be regarded as a noncommutative spacetime model which cannot be deduced from a doubly-quantized phase space, since it cannot be obtained as the $\hbar\to0$ limit of such a phase space. Conversely, the $\lambda$-Minkowski noncommutative spacetime model can be embedded into a doubly quantized picture, and deduced from there in the smooth $\hbar\to 0$ limit.

In this work we considered only models with pure spatial noncommutativity. The issue of double quantization in models with a noncommuting time coordinate seems to be nontrivial. 
In fact, 
for those models for which a nontrivial structure regarding time is present, such as the $\kappa$-Minkowski model, the Moyal twist of quantum mechanics is ill-defined, since in it only the vector fields of the space components appear.  Given the relevance of models with time noncommutativity in the literature, especially when phenomenology is involved \cite{Addazi:2021xuf, Amelino-Camelia:2008aez}, it would be interesting to investigate whether other methods leading to double quantization can be defined, which are compatible with a time noncommutativity.

A possible way forward in this direction could start from results on the pregeometry of $\kappa$-Minkowski \cite{Amelino-Camelia:2012vlk}, based on a covariant formulation of quantum mechanics \cite{Reisenberger:2001pk}. In \cite{Amelino-Camelia:2012vlk} the pregeometric prescription was aimed at giving a firmer ground to the construction of the kinematical and physical Hilbert space of the states describing points in $\kappa$-Minkowski spacetime. The authors of \cite{Amelino-Camelia:2012vlk} did not  consider the effects of the phase-space noncommutativity parameter $\hbar$, which was set to $1$. Reintroducing explicitly $\hbar$, and working out the full phase space algebra, one gets (we use here the same notation as in \cite{Amelino-Camelia:2012vlk})
\be
{[}x_i,  x_0{]}\,=\,  i\hbar \ell x_i\, , \qquad {[}x_i,  x_j{]}\,=0 \, ,\qquad {[}x_i,  \pi_j{]} \,=\, i\hbar e^{\ell \pi_0}\,, \qquad
{[} x_0,  \pi_0{]} \,=\, -i\hbar\,, \qquad {[} \pi_\alpha,  \pi_\beta{]} \,=\, 0\,.\label{kM}
\ee

Taking this at face value, we see that in this case one does recover the usual commutation relations of quantum mechanics in the $\ell\to0$ limit. The $\hbar\to 0$ limit can be made sense of in the same way as done in standard quantum mechanics, where the commutators reduce to Poisson brackets as $\frac{1}{i\hbar} [\cdot,\cdot]\xrightarrow{\hbar \to 0} \{\cdot,\cdot\}$. In contrast with the $\lambda$-Minkowski model, notice that  pure spacetime noncommutativity is not allowed in this construction, but nontrivial spacetime Poisson brackets are.  A similar feature is found in the Snyder model \cite{snyder}. In \cite{Ballesteros:2019mxi, Gubitosi:2020znn} the full phase-space algebra was derived, including explicitly both $\hbar$ and the spacetime noncommutativity parameter $\Lambda$ (which in that case is just the inverse of the momentum space curvature):
\be
\begin{array}{ll}
\left[ x^0,  x^i\right]\, =\,   i\hbar\,\Lambda \,   K_{i} \, ,& \quad \left[  x^i,  x^j\right]\, =\,  i\hbar\, \Lambda\,\epsilon_{ijk}   J_k\, ,  \\[4pt]
{ \left[ x^0, p_{\alpha}\right]\, =\,  i\hbar\,(\delta_{0\alpha}-\frac{\Lambda}{c^2} \,  p_{0} p_{\alpha})\, ,}& \quad {  \left[ x^i, p_{j}\right]\, =\,  i\hbar\,(\delta_{ij}+\Lambda \,  p_{i} p_{j})\, , }
\\[4pt]
\left[ x^i, p_{0}\right]\, =\,  i\hbar\,\Lambda \,  p_{0} p_{i}\, ,  &\quad  \left[ p_{\alpha}, p_{\beta}\right]\, =\, 0\, .
\end{array}
\label{Snyder}
\ee
Again, the $\hbar\to 0$ limit can be performed in such a way that all commutators turn into Poisson brackets, including the ones between spacetime coordinates, while the $\Lambda\to 0$ limit reproduces standard quantum mechanics.
Summarizing, the difference between these cases with noncommutative time coordinate and the one of $\lambda$-Minkowski with pure spatial noncommutativity is that, in \eqref{kM} and \eqref{Snyder}, the $\hbar\to 0$ limit necessarily turns the phase-space and spacetime commutators into Poisson brackets (pure spacetime noncommutativity is not possible, but only deformed Poisson brackets), while in our $\lambda$-Minkowski doubly quantized phase space, Eq.~\eqref{commrel3}, one sees that pure spatial noncommutativity emerges as the  $\hbar\to 0$ limit of the doubly quantized phase space.\footnote{Also notice the these models are invariant under (deformed) Poincar\'e symmetries. In order to make the comparison to the results of this work more direct, one should look at the Galilean limit of these models,  invariant under (deformed) Galilei symmetries \cite{Ballesteros:2020uxp, Ballesteros:2019mxi, Ballesteros:2021dob}}

It would also be interesting to perform the analysis at the level of operator representation of the commutation relations, along the lines of~\cite{Lizzi:2018qaf, Lizzi:2019wto}, and see how the various limits affect the possibility to localize the states. Work along these lines is in progress.

Another recently-developed direction of investigation which might lead to a better understanding of the links between quantum properties of spacetime and quantum-mechanical phase-space noncommutativity is that focusing on  ``quantum reference frames''~\cite{Giacomini:2017zju}. In this construction, reference frames in quantum mechanics are associated to a quantum system rather than to the  classical laboratory, so they acquire  quantum properties and the commutative parameters of standard Galilean transformations are turned into quantum operators. If spacetime is defined in terms of such quantum reference frames, then it inherits their quantum properties. While this statement still needs to be made more precise, and work on this is in progress, recent preliminary results showed that the associated symmetry group defining transformations between quantum reference frames is a more general group than that of Galilei symmetries~\cite{Ballesteros:2020lgl}.

\subsection*{Acknowledgements}
We acknowledge support from the INFN Iniziativa Specifica GeoSymQFT and INFN Iniziativa Specifica QUAGRAP.  The authors would also like to thank support from the COST Action CA18108.  FL~acknowledges financial support from the State Agency for Research of the Spanish Ministry of Science
and Innovation through the ``Unit of Excellence Maria de Maeztu 2020--2023'' award to the Institute of Cosmos Sciences (CEX2019-000918-M) and from PID2019-105614GB-C21 and
2017-SGR-929 grants.

\bibliography{ref}
\bibliographystyle{utphys}
\end{document}